Original Paper

# Exploring Longitudinal Cough, Breath, and Voice Data for COVID-19 Progression Prediction via Sequential Deep Learning: Model Development and Validation

Ting Dang[1], PhD; Jing Han[1*], PhD; Tong Xia[1*], MPhil; Dimitris Spathis[1], PhD; Erika Bondareva[1], MRES; Chloë Siegele-Brown[1], PhD; Jagmohan Chauhan[1,2], PhD; Andreas Grammenos[1], PhD; Apinan Hasthanasombat[1], MPhil; R Andres Floto[1], PhD; Pietro Cicuta[1], PhD; Cecilia Mascolo[1], PhD

[1]Department of Computer Science and Technology, University of Cambridge, Cambridge, United Kingdom
[2]Electronics and Computer Science, University of Southampton, Southampton, United Kingdom
*these authors contributed equally

**Corresponding Author:**
Ting Dang, PhD
Department of Computer Science and Technology
University of Cambridge
15 JJ Thomson Ave
Cambridge, CB3 0FD
United Kingdom
Phone: 44 7895587796
Email: td464@cam.ac.uk

## Abstract

**Background:** Recent work has shown the potential of using audio data (eg, cough, breathing, and voice) in the screening for COVID-19. However, these approaches only focus on one-off detection and detect the infection, given the current audio sample, but do not monitor disease progression in COVID-19. Limited exploration has been put forward to continuously monitor COVID-19 progression, especially recovery, through longitudinal audio data. Tracking disease progression characteristics and patterns of recovery could bring insights and lead to more timely treatment or treatment adjustment, as well as better resource management in health care systems.

**Objective:** The primary objective of this study is to explore the potential of longitudinal audio samples over time for COVID-19 progression prediction and, especially, recovery trend prediction using sequential deep learning techniques.

**Methods:** Crowdsourced respiratory audio data, including breathing, cough, and voice samples, from 212 individuals over 5-385 days were analyzed, alongside their self-reported COVID-19 test results. We developed and validated a deep learning–enabled tracking tool using gated recurrent units (GRUs) to detect COVID-19 progression by exploring the audio dynamics of the individuals' historical audio biomarkers. The investigation comprised 2 parts: (1) COVID-19 detection in terms of positive and negative (healthy) tests using sequential audio signals, which was primarily assessed in terms of the area under the receiver operating characteristic curve (AUROC), sensitivity, and specificity, with 95% CIs, and (2) longitudinal disease progression prediction over time in terms of probability of positive tests, which was evaluated using the correlation between the predicted probability trajectory and self-reported labels.

**Results:** We first explored the benefits of capturing longitudinal dynamics of audio biomarkers for COVID-19 detection. The strong performance, yielding an AUROC of 0.79, a sensitivity of 0.75, and a specificity of 0.71 supported the effectiveness of the approach compared to methods that do not leverage longitudinal dynamics. We further examined the predicted disease progression trajectory, which displayed high consistency with longitudinal test results with a correlation of 0.75 in the test cohort and 0.86 in a subset of the test cohort with 12 (57.1%) of 21 COVID-19–positive participants who reported disease recovery. Our findings suggest that monitoring COVID-19 evolution via longitudinal audio data has potential in the tracking of individuals' disease progression and recovery.

**Conclusions:** An audio-based COVID-19 progression monitoring system was developed using deep learning techniques, with strong performance showing high consistency between the predicted trajectory and the test results over time, especially for recovery trend predictions. This has good potential in the postpeak and postpandemic era that can help guide medical treatment and optimize hospital resource allocations. The changes in longitudinal audio samples, referred to as audio dynamics, are associated with COVID-19 progression; thus, modeling the audio dynamics can potentially capture the underlying disease progression process and further aid COVID-19 progression prediction. This framework provides a flexible, affordable, and timely tool for COVID-19 tracking, and more importantly, it also provides a proof of concept of how telemonitoring could be applicable to respiratory diseases monitoring, in general.

# Introduction

## Background

Since the beginning of the SARS-CoV-2 pandemic in January 2020, different methods have been developed and used for diagnostic testing and screening. In addition to the most commonly adopted laboratory tests via reverse transcription polymerase chain reaction (RT-PCR) [1,2] or chest computed tomography (CT) scans [3] for diagnosis, a variety of digital technologies, often using artificial intelligence, have also been investigated for COVID-19 screening [4-7]. Among these, automatic audio-based COVID-19 detection has drawn increasing attention due to its numerous advantages, including its flexible, affordable, scalable, noninvasive, and sustainable data collection methods.

The existing literature has mainly investigated the information content of different audio modalities (eg, cough, breathing, and voice) [8-12] and the power of various machine learning techniques, especially deep learning for COVID-19 detection [10,13-19]. Although success has been witnessed recently in COVID-19 detection from audio signals through machine learning techniques [10], there is still a paucity of work on continuous monitoring of COVID-19 progression. This could provide individual-specific, timely indication of disease development at scale, guide personalized medical treatments, potentially capture disease onset to curb transmission, and estimate the recovery rate, which plays a key role in determining quarantine rules during the current postpeak and postpandemic times. It would also allow better resource management in health care systems, while remotely monitoring patients and bringing them to the hospital (only) when necessary. Evidence suggests that COVID-19 progression varies among individuals, with the mean disease duration ranging from 11 to 21 days, depending on gender and age, comorbidities, variants of SARS-CoV-2, and time receiving treatment [20-25]. By continuously monitoring patients' disease progression, individual-specific information could be captured to benefit both patients and doctors. In addition, compared to the commonly adopted uncomfortable diagnostic methods of RT-PCR tests and radiation-intensive CT scans that are conducted on hospital sites, audio-based monitoring of disease progression can be nonintrusively repeated on a daily basis and for prolonged periods, proving a good fit for longitudinal remote monitoring.

Recent work has recognized the use of a 3-escalating-phase description of COVID-19 progression [26], including early infection, pulmonary involvement, and systemic hyperinflammation, which demonstrates commonality in disease progression. We hypothesize that this could be captured longitudinally via audio signals in automatic monitoring systems. Though the participants in our study may not experience all 3 stages, it is assumed that audio characteristics are affected during clinical progression of the disease. Figure 1 shows spectrograms of 1 participant reading the same sentence over a 43-day period who reported COVID-19 infection followed by recovery. As indicated in the black box, the fundamental frequency and its harmonics were not clearly separable when the participant tested positive (top row), especially on November 14, 2020, indicating a lack of control of vibration of the vocal cords. The separability increased after recovery. This matches the observed clinical course of COVID-19 progression [22], with the least separability 5 days after the first positive test result (November 14, 2020) and an increase in separability 9 days after infection (November 18, 2020). Similar patterns were also observed for the harmonics in the high-frequency range from 2 to 4 kHz (blue box). There was no obvious difference in the spectrogram patterns between November 18 (positive) and November 22 (negative), reflecting the difficulty of the COVID-19 detection task in general. Overall, this evolution of the spectrograms with disease progression shows that COVID-19 infection can manifest as changes in acoustic representations. As disease progression varies among individuals (eg, different recovery times or different severity levels), longitudinal audio changes could vary among individuals. However, it is common that longitudinal audio changes present with COVID-19, and modeling them can potentially benefit COVID-19 progression prediction.

**Figure 1.** Sound recordings have distinct features during disease progression. This is evident here in the spectrograms of 1 participant who repeated the same sentence on 6 different days. The participant reported positive test results from November 10 to 18, 2020, and reported negative test results from November 22 to December 26, 2020, indicating a recovery trend. The fundamental frequency and its harmonics (black box) for positive recordings demonstrate a lack of control in vocal cords, indicated by their nonseparability. An increasing separability can be seen from positive to negative recordings over time, suggesting the recovery of voice characteristics. Similarly, the harmonics in the frequency range (2-4 kHz, blue box) manifest increasing separability, also reflecting the recovery trend.

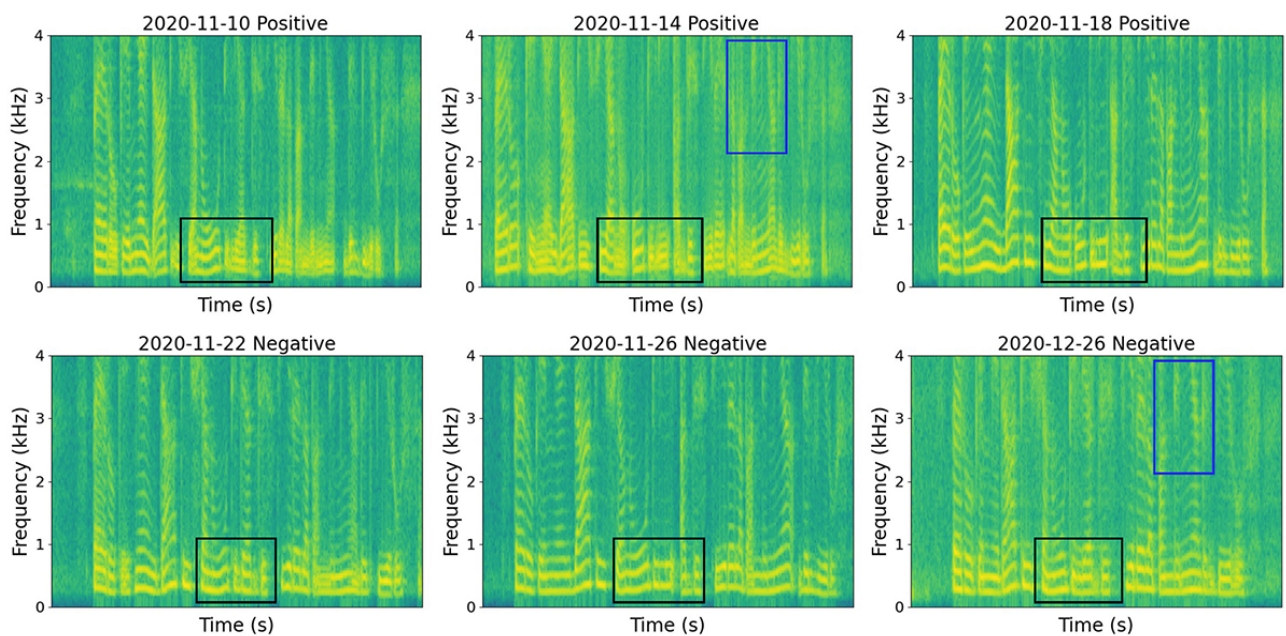

Furthermore, audio characteristics vary among individuals (eg, one COVID-19–*positive* participant may produce a similar spectrogram as another COVID-19–*negative* participant). This is not considered in most conventional audio-based COVID-19 detection systems, which so far have only used a single audio sample rather than sequences. This makes automatic detection a difficult task and may lead to wrong predictions. While longitudinally modeling the evolution of spectrograms, the individual's past audio signal can serve as a baseline and the predictions can be corrected, given this reference. Additionally, the spectrogram for each individual when healthy can be used as a reference for its own infected status, and longitudinally modeling the relative changes in the audio sequences is likely to be more accurate for COVID-19 detection. In a much generalized sense, the mean and SD of the audio recordings in an individual's healthy state are both personalized. This provides a good threshold for the nonhealthy states and benefits COVID-19 detection. Motivated by these advantages, we explored the potential of longitudinally modeling sequential audio recordings as biomarkers of disease progression, focusing on how best to capture dynamic changes in the audio sequences over time and aiming to demonstrate predictive power.

## Objective

In this study, we developed an audio-based COVID-19 progression prediction model using longitudinal audio data. We adopted sequential deep learning models to capture longitudinal audio dynamics and to make predictions of disease progression over time. First, we examined whether modeling audio dynamics could aid in COVID-19 detection. This showed strong performance compared to conventional models using a single audio sample. We then evaluated our model's performance in predicting disease progression trajectories: our predictions successfully tracked test results and also matched the statistical analysis in the timeline and duration of COVID-19 progression. In particular, we explored the use of audio signals for recovery prediction, as this may be useful in relation to setting home quarantine requirements. From a public health perspective, an approach such as the one we propose has potential implications for how infected individuals are monitored, namely it could allow more fine-grained remote tracking and hence more efficient management of health system resources by keeping individuals out of the hospital as much as possible.

## Methods

### Study Design and Overview

We investigated whether longitudinally modeling audio biomarkers (cough, breath, and voice) can benefit COVID-19 detection and whether it could be used to monitor disease progression accurately and in a timely manner (Figure 2). The audio sequences were modeled by recurrent neural networks with gated recurrent units (GRUs) to consider audio dynamics, which reflect disease progression. The investigation was divided into 2 subtasks: one concerning COVID-19 detection by predicting audio biomarkers as positive and negative and the other concerning disease progression trajectory monitoring, examining the predicted probability of being positive over time. For instance, a decrease in the probability of being positive over time indicates a recovery trend, while an increase indicates a worsening trend. The first subtask aimed to assess whether modeling past audio biomarkers in the input space benefits COVID-19 detection in general, while the second subtask focused on longitudinal analysis of disease progression in the output space.

**Figure 2.** Overview of study design: COVID-19 detection and progression were assessed from audio data. Voice, cough, and breathing sound recordings were collected from each participant over a period, together with self-reported COVID-19 test results. During model development, audio recordings were chunked into segments consisting of 5 samples covering a few days and processed using sequential modeling techniques (GRUs) for COVID-19 monitoring. Two subtasks were evaluated: (1) COVID-19 detection (positive vs negative) and (2) disease progression monitoring. GRU: gated recurrent unit.

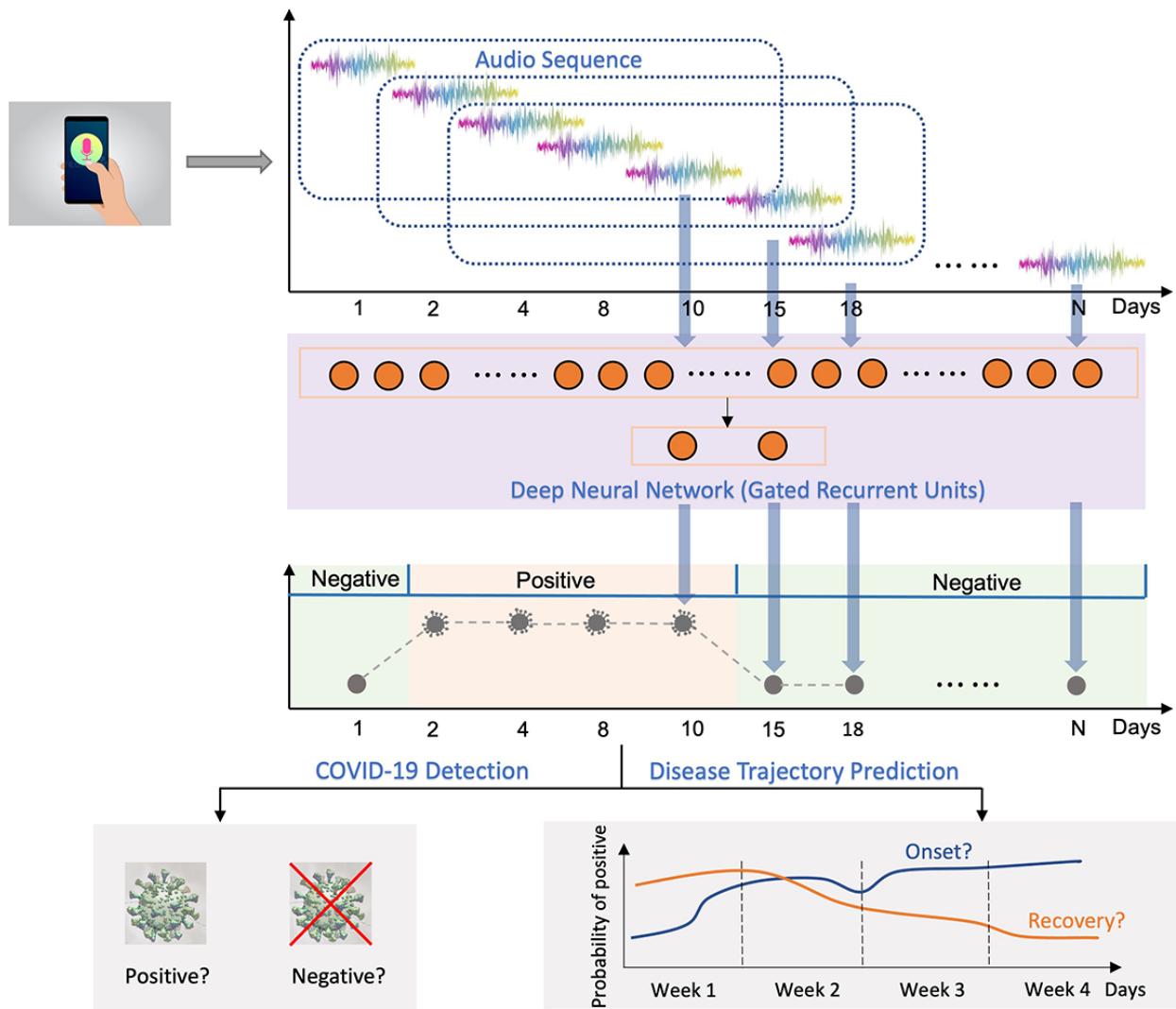

### Data Set Preparation and Statistics

A mobile app [27] was developed and released in April 2020 for data gathering, aiming to crowdsource participants' audio recordings, COVID-19 test results, and demographics, medical history, and COVID-19–related symptoms. Each participant was asked to record 3 different audio sounds on a single day, including a cough recording with 3 voluntary cough sounds, a breathing recording with 3-5 breathing sounds, and a voice recording where each participant was asked to read a short phrase displayed on the screen, 3 times. The COVID-19 test results were self-reported, chosen from a positive report, a negative one, or not having been tested. No specific methods for the diagnosis were required, which can be a lateral flow test result, an RT-PCR test result, or a CT scan result. Participants were encouraged to provide data regularly. More details can be found in Multimedia Appendix 1 and Xia et al [28].

From April 2020 to April 2021 (Figure 3a), 3845 healthy participants (COVID-19–negative) and 1456 COVID-19–positive participants contributed audio samples with positive or negative clinical self-reported test results for at least 1 day. We used these participants' data if 5 or more samples were provided, resulting in 447 (11.6%) and 168 (11.5%) negatively and positively tested participants, respectively. Label quality was manually checked to remove unreliable users, and audio recording quality was examined using Yet Another Mobile Network (YAMNet) [29] to filter out corrupted and noisy samples, leaving 106 (63.1%) COVID-19–positive participants. To generate a balanced data set, a cohort of 212 longitudinal users in total (106, 50%, COVID-19–positive and 106, 50%, COVID-19–negative) was selected across different countries.

**Figure 3.** Data flow diagram and demographic statistics. Large data sets were required to identify and avoid biases. (a) Data selection process. (b) Demographic statistics of eligible participants, including language, gender, age, and symptoms. English was the dominant language, comprising 54.2% (n=115) of the cohort. Age and gender were relatively balanced between positive and negative groups. In addition, 100 (94.3%) COVID-19–positive participants and 82 (77.4%) COVID-19–negative participants reported COVID-19 symptoms. (c) Duration and reporting intervals in terms of days and number of samples. The median number of samples was 9 (left), corresponding to a time span of 35 days (middle left). COVID-19–negative participants reported for a longer period compared to COVID-19–positive participants. The median reporting interval for the cohort was 3 days (middle right), validating the effective temporal dependencies of the audio data. The median duration after augmentation was 17 and 18 days for COVID-19–positive and COVID-19–negative participants, respectively (right), showing that the augmentation eliminated the confounding effects of the different duration for the 2 subgroups.

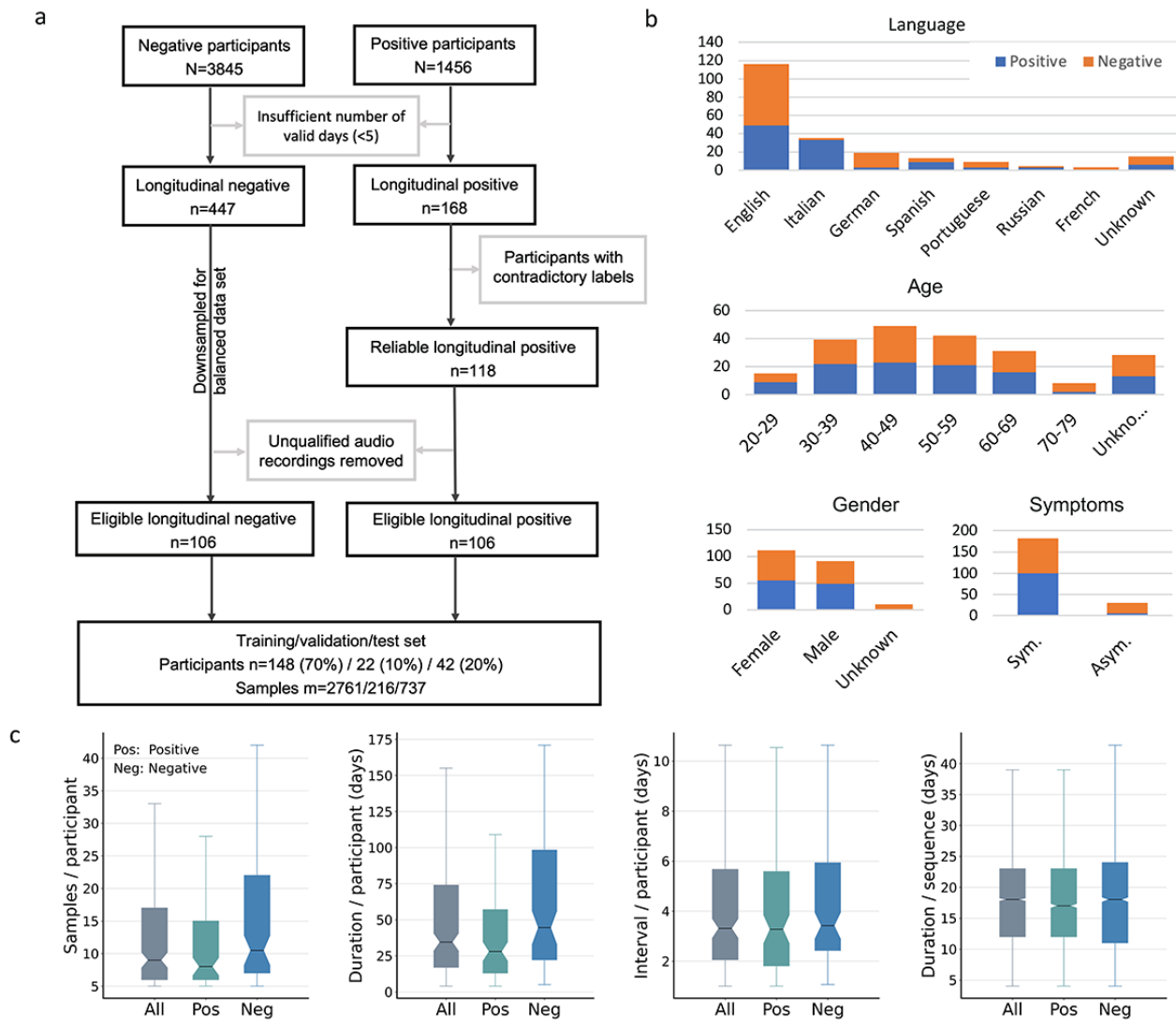

The mobile app is a multilanguage platform, and the cohort consisted of 8 different languages, with 115 (54.2%) English users as the dominant subgroup (Figure 3b). Age and gender were relatively balanced between COVID-19–positive and COVID-19–negative groups, with 110 (51.9%) female participants and 142 (67.0%) participants aged between 30 and 59 years. In addition, 100 (94.3%) of 106 COVID-19–positive participants and 82 (77.4%) of 106 COVID-19–negative participants reported COVID-19 symptoms, such as loss of taste or smell, and fever. The median number of samples for each user was 9 (Figure 3c, left), corresponding to a period of 35 days (Figure 3c, middle left). The median duration for COVID-19–positive participants was smaller than for COVID-19–negative participants, namely 28 and 45 days, respectively. This duration is expected to contain adequate audio dynamics associated with disease progression and to cover the complete course of disease progression for most participants [21]. In addition, the reporting interval for each participant was also computed as the average of the time intervals between 2 consecutive samples, and the median value was 3 days for both COVID-19–positive and COVID-19–negative groups (Figure 3c, middle right), validating the temporal dependencies of the data. To develop machine learning models, data augmentation was carried out, and the duration after augmentation for COVID-19–positive and COVID-19–negative participants was balanced, with a median value of 17 and 18 days, respectively (Figure 3c, right). This duration aligns with the disease progression duration that is generally from 11-21 days [20-24]. The similar duration for COVID-19–positive and COVID-19–negative participants after augmentation also helped to eliminate the confounding effect of the original different duration for the 2 groups in model development (Figure 3c, left). The data were split into training, validation, and test sets, with 148 (70%), 22 (10%), and 42 (20%) balanced COVID-19–positive and COVID-19–negative participants, respectively, as well as the relatively balanced languages and genders (see Multimedia Appendix 2).

## Data Processing

To effectively develop the deep learning model, we ensured a sufficient amount of appropriately processed data available for modeling by performing audio preprocessing, sequence generation, and data augmentation.

### Audio Preprocessing

Audio recordings were first resampled to 16 kHz and converted to a mono channel. These audio recordings were then preprocessed by removing the silence periods at the beginning and end of the recordings, normalized to a maximum amplitude of 1.

### Sequence Generation

To increase the number of audio sequences for model development, the audio samples for each participant were chunked into short sequences with a fixed number of 5 samples. To ensure the 5 samples within an audio sequence contained effective and adequate audio dynamics in COVID-19 progression, a further constraint was applied, limiting the maximum time gap between 2 subsequent samples to 14 days. Any sequences that violated this constraint were removed. This resulted in sequence lengths ranging from 5 to 56 days, covering the time span of disease progression.

### Data Augmentation

Though the data set was selected to balance the number of participants for COVID-19–positive and COVID-19–negative groups, the number of samples for each participant was still different (refer to Figure 3c). The COVID-19–negative participants provided more samples than COVID-19–positive participants, and the COVID-19–positive participants also provided negative samples after recovery, leading to a significantly larger number of COVID-19–negative samples than COVID-19–positive samples in the cohort. Further, the data set was relatively small. Therefore, 3 augmentation techniques were used to increase the data size and to balance the COVID-19–positive and COVID-19–negative samples (see Multimedia Appendix 1).

## Model Architecture

The proposed model consisted of a pretrained network of VGGish (Google) for feature extraction and a recurrent neural network of GRUs for disease prediction (Figure 4). Three different modalities (breathing, cough, and voice recordings) were adopted as the input. For each modality, audio recordings were first converted to spectrograms and then fed into a VGGish pretrained network for higher-level feature representation, which could help leverage and transfer the knowledge learned from an external massive general audio data set [29]. The embeddings converted by VGGish from the 3 modalities were concatenated to form a multimodal input vector for the subsequent GRU-based prediction network. The reason for choosing GRUs over a long short-term memory neural network is fewer parameters in limited data size regimes. GRUs also use less memory and execute faster, which would be a benefit during potential model deployment. The outputs from the GRUs were evaluated for 2 different tasks: one estimating the model capability in binary diagnosis by taking the binary output of the model and the other predicting the disease progression trajectory by utilizing the probabilities of positive predictions. Further details can be found in Multimedia Appendix 1.

**Figure 4.** Model structure. A pretrained convolutional neural network (CNN)–based model VGGish was used as the feature extractor, and GRUs were used as a classifier, followed by dense layers, to account for longitudinal audio dynamics. This is a multitask learning framework, with COVID-19 detection as the main task and language detection as an auxiliary task to avoid language bias. $h_i$, i ∑ [1,2,…N] represents the hidden vectors in the GRUs for time step $t_i$. The reverse layer is used for the language task, as shown in Multimedia Appendix 1, Equation (5). GRU: gated recurrent unit.

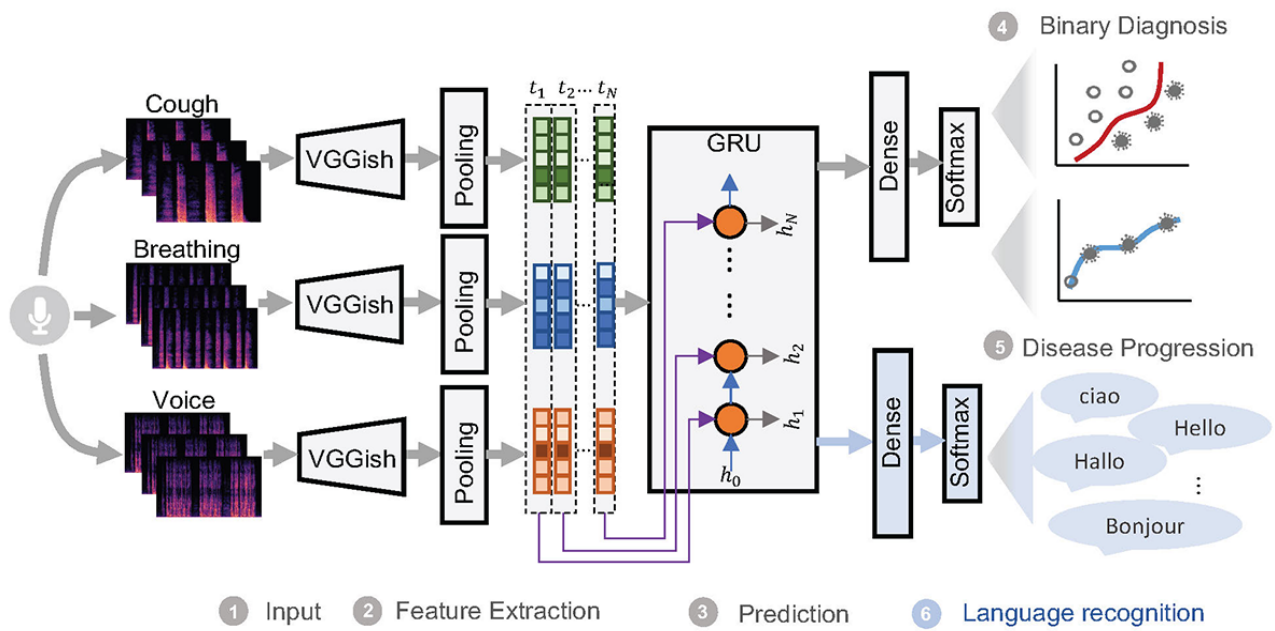

Since the data set contained voices in 8 different languages and the disease prevalence for each language was different (Figure 3b), a language bias may have been introduced in the machine learning models, leading to the model recognizing the language instead of COVID-19–related information (eg, classifying Italian speakers as COVID-19–positive and English speakers as COVID-19–negative owing to the higher prevalence of the disease in Italian-speaking users and lower prevalence in English speakers). To reduce the language impact, a multitask learning framework was proposed to include the auxiliary task of language recognition simultaneously with the COVID-19 prediction task (see Multimedia Appendix 1).

## Performance Evaluation

As the sequence length in the training data varied within the range of 5-56 days, the model was able to capture longitudinal audio dynamics with varying length. Therefore, during the inference stage, the prediction was made, given all the past audio recordings with no fixed number of samples. This was slightly different from the training phase but was more practical in real applications. To maintain the effective temporal dependencies of the audio recordings and match the training, a time constraint was applied to account only for the past audio recordings within 56 days before the current day, the maximum duration in the training phase. Further, we evaluated the model performance from the second sample of each participant to make sure the predictions captured the longitudinal audio dynamics.

### COVID-19 Detection

For COVID-19 detection, the performance was measured using the AUROC, sensitivity, and specificity. The AUROC illustrated the diagnostic ability of the binary classifier. Sensitivity showed the model's ability to identify correctly COVID-19–positive samples, while specificity showed the model's ability to correctly identify samples without the disease.

### Disease Progression Prediction

For the disease trajectory prediction, model performance was evaluated for each participant. There were 2 different metrics used based on individuals' test labels. For participants who recorded any transitions between positive and negative test results during the reported period, we adopted the point-biserial correlation coefficient $\gamma_{pb}$ to measure the correlation between the predicted probability of positive test results and test labels. For participants who reported positive and negative test results consistently over the reported period, it was not possible to compute the correlation between continuous predictions and test labels. Therefore, we adopted the accuracy $\gamma$ computed as the ratio of the correctly predicted samples over the total number of samples (see Multimedia Appendix 1). Though $\gamma_{pb}$ ranges between –1 and 1 and $\gamma$ is in the range of 0-1, the higher the value, the better the predictions for both metrics. It is also expected that $\gamma_{pb}$ achieves a positive correlation from 0 to 1 for a good model. Therefore, we reported performance by combining these 2 measures of $\gamma_{pb}$ and $\gamma$.

### Recovery Trajectory Prediction

We further examined model performance for the recovery subgroup in the test cohort, where 12 (57.1%) of 21 COVID-19–positive participants reported a recovery trend in their test results. We adopted $\gamma_{pb}$ as the evaluation metric. As we noted that there may be a delay in participants taking clinical tests or reporting results, negative predictions could be earlier than self-reported negative test results, which is acceptable. We further aligned predictions and test results temporally using dynamic time warping (DTW) and computed $\gamma_{pb}$ with aligned predictions to account for the delay.

### Latent Space Visualization

To gain an in-depth understanding of the model, we aimed to compare the intermediate audio representations of the model for different participants, including a participant who reported infection followed by recovery, a participant who continuously reported positive test results, and a healthy participant. The intermediate audio representations were taken as latent vectors from the last hidden layer of the model and further projected to a 4-dimension latent space using principal component analysis.

## Statistical Analysis

### Disease Progression With Symptoms

We studied the correlations between the predicted probability and symptoms for COVID-19–positive and COVID-19–negative participants, respectively. For COVID-19–positive participants, we assumed that the number of symptoms is correlated with the severity of illness; thus, a high probability of positive predictions was expected for the audio recordings reported alongside more symptoms. This led to a high correlation between the predicted probability and the number of symptoms. On the contrary, for the healthy participants, the number of symptoms was not correlated with the severity of illness; thus, no correlation was expected. This can validate whether the model is capable of learning COVID-19–related information instead of symptom-related information.

### Disease Progression in the First 7 Days

Evidence on chest X-ray or CT showed that 56 (22.6%) patients presenting with disease experienced resolution 7 days after symptom onset, 30 (12.1%) showed a stable condition, and the remaining 162 (65.3%) patients worsened within 7 days from symptom onset [26]. We analyzed the predicted trajectories over a similar period, namely the first 7 days, to compare the statistics of the predicted trend with the reported ones. Though many participants reported symptoms on the first day they started recording, which may not be the first day they experienced symptoms, the increasing trend in the first few days could still suggest initial worsening of the patients' condition. We defined the 7-day window to be from either the first day of reported symptoms or the day of the first positive test if no symptoms were reported before that.

## Ethical Considerations

The study was approved by the ethics committee of the Department of Computer Science at the University of Cambridge (with ID #722). Our mobile app displays a consent screen, where we ask the user's permission to participate in the study by using the app.

## Results

### COVID-19 Detection

To determine whether considering audio dynamics via the sequential modeling techniques of GRUs is effective in detecting COVID-19, the performance was compared against 2 benchmarks that do not capture audio dynamics (Figure 5): one uses only audio biomarkers of the same day for prediction, while the other uses the average feature representation of audio sequences in the prior days for the last day (Figure 5a). The proposed system (Figure 5b) outperforms the 2 benchmarks with the highest AUROC of 0.79 (0.74-0.84), a sensitivity of 0.75 (0.67-0.82), and a specificity of 0.71 (0.67-0.75), yielding 19.7% and 31.7% relative improvement in terms of the AUROC over the 2 benchmarks and demonstrating the effectiveness of modeling longitudinal audio biomarkers. We used the 1-tailed $z$ test to validate the significance of the performance improvement in terms of the AUROC of the proposed approach over 2 baselines. We found that the proposed approach significantly improved the performance over "single" ($P=.09$) and "average" ($P=.03$).

**Figure 5.** The proposed sequential model shows superior performance in COVID-19 detection compared to benchmarks leveraging only 1 isolated audio data point per user. (a) "Average" means using the average of feature representations within the sequence for prediction, and "Single" means using only the feature representation on the same day for prediction. None of these systems capture longitudinal voice dynamics. (b) The proposed sequential modeling outperformed 2 benchmarks, suggesting the advantages of capturing disease progression via voice dynamics. (c) Individual-level accuracy for 42 participants in the test cohort.

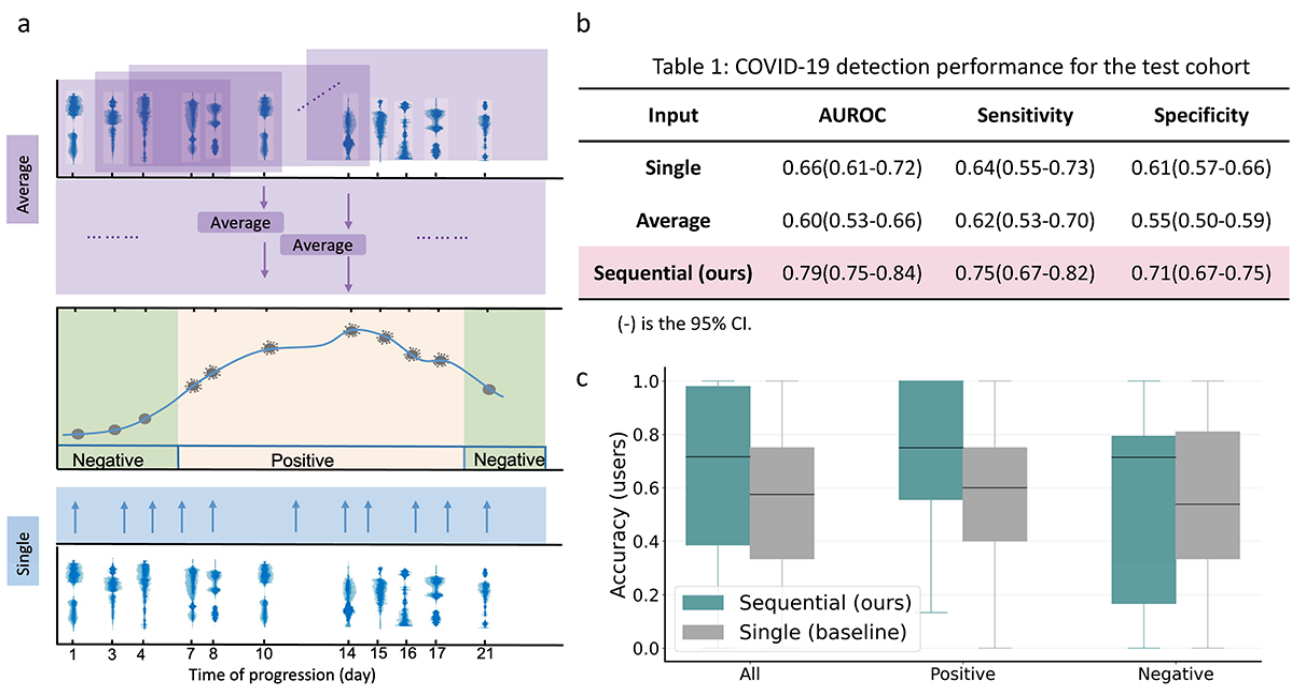

For further assessment of whether sequential modeling using past audio biomarkers could provide an adjustable baseline for each individual, the prediction accuracy for each participant was evaluated and compared to the "single" benchmark, defined as the ratio of correctly predicted samples over the total number of samples for each participant (Figure 5c). We observed that the proposed sequential modeling outperformed the "single" benchmark. The performance range of the sequential model for negative participants was larger than the benchmark, due to worse performance on 2 individuals.

### Disease Progression Prediction

We analyzed the predicted disease progression trajectory by comparing it with the test results. The predicted progression trajectory is represented by the probability of positive prediction within the range of 0-1 over time, with a higher value indicating a high possibility of positive test results (Figure 6). Three different disease progression trajectories are shown in Figure 6a, Figure 6b, and Figure 6c, respectively. For the recovering participant P1 (Figure 6a), a high probability was observed when P1 tested positive and a low probability when P1 tested negative. The model performance was evaluated using the point-biserial correlation coefficients $\gamma_{pb}$ that measured the correlation between the predicted trajectory and the test results. Our model achieved $\gamma_{pb}=0.86$ for P1, demonstrating a strong capability to predict disease progression. We further categorized the probability over 0.5 as a positive prediction and below 0.5 as a negative prediction. The predictions also matched the test results.

**Figure 6.** Our approach enabled prediction of disease progression. Orange and cyan indicate positive and negative test results, and + and • represent positive and negative predictions, respectively. The green star indicates the presence of symptoms. (a) Disease progression of recovering participant P1. (b) Disease progression of COVID-19–positive participant P2. (c) Disease progression of COVID-19–negative participant P3. (d) Overall performance for the test cohort in terms of the point-biserial correlation coefficient $\gamma_{pb}$/accuracy $\gamma$.

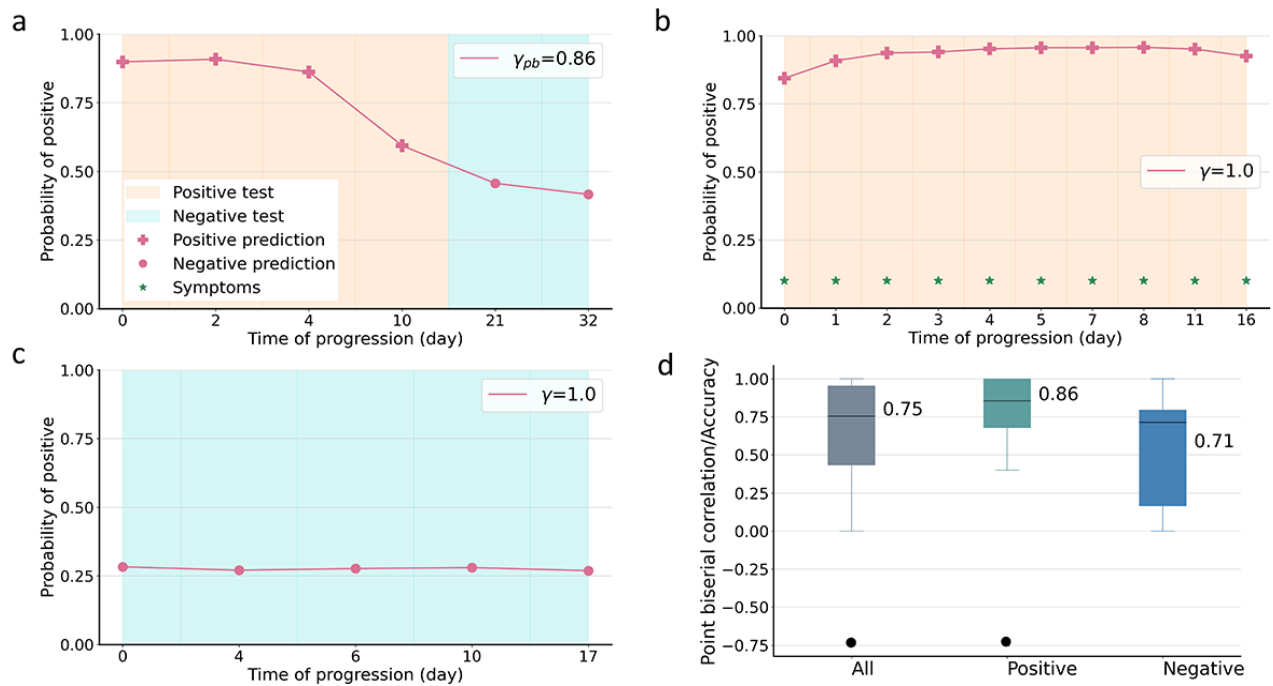

For the COVID-19–positive participant P2 who consistently reported positive test results (Figure 6b), our model outputs positive predictions matching the test results. It should be noted that $\gamma_{pb}$ is not applicable to participants who report consistently positive or negative test results; therefore, accuracy $\gamma$ was instead used, which computes the ratio of correct predictions in terms of positive or negative predictions over the total number of samples. Participant P2 had $\gamma=1$ since all the predictions were positive. Further, the probability of positive predictions increased from symptom onset day 0 to day 8 and decreased slightly to day 16, which matches the clinical course in general.

For the COVID-19–negative participant P3 (Figure 6c), the predicted probability was consistently below 0.5, corresponding to negative predictions that align with the test results. One type of disease progression that transits from negative to positive was not included due to the limited number of participants in the cohort. Though we adopted time-inverse augmentation (see Multimedia Appendix 1) that reverses the audio biomarkers and their corresponding labels in time to enrich the different disease progressions, especially the negative-to-positive transitions, the time-reversed audio biomarkers and disease progression may still not match the actual progression and cannot be well captured in the model.

Figure 6d demonstrates the model performance in disease progression monitoring for the entire test cohort, where $\gamma_{pb}$ and $\gamma$ were adopted for different participants. Our model achieved 0.75 for all the 42 (100%) test participants and 0.86 and 0.71 for COVID-19–positive and COVID-19–negative participants, respectively. Some more examples are given in Multimedia Appendix 3 and Multimedia Appendix 4.

### Recovery Trajectory Prediction

The predicted recovery trajectory of 2 randomly selected participants P4 and P5 are presented in Figure 7a and Figure 7b. For participant P4, a slight increase in probability was observed from day 0 to day 2, suggesting an increase in the severity of illness. The predicted probability decreased from day 2, showing a recovery trend. The categorized positive and negative predictions also matched the test results, except for day 27, with our model predicting negative result and the test result still being positive. This is possibly due to a delay in participants taking clinical tests or reporting results; therefore, earlier negative predictions are acceptable. This also suggests the advantages of our audio-based data, which are precisely timed and can be instantly analyzed.

For a weakly predicted trajectory of participant P5 in Figure 7b, a predicted recovery trend with decreasing probability was clearly observed. However, the probabilities were all categorized as positive predictions even after the user tested negative from day 16 to day 23. This is possibly due to (1) individual differences in audio characteristics or (2) an asymptomatic participant exhibiting minor changes in audio characteristics, thus leading to a slowly recovering trend prediction.

The overall performance for all 12 (100%) recovering participants is reported in Figure 7c, with $\gamma_{pb}=0.76$. As negative predictions with a time shift from negative test results are acceptable (as shown in Figure 7a), we aligned predictions and test results temporally using DTW. The model further achieved $\gamma_{pb}=0.86$, demonstrating a strong capability to monitor recovery. Some more examples can be found in Multimedia Appendix 3 and Multimedia Appendix 4.

**Figure 7.** Recovery trends can be predicted. The orange and cyan areas indicate positive and negative test results, respectively. The predictions above 0.5 were categorized as positive predictions (+) and below 0.5 as negative predictions (•). (a,b) Recovery predictions for 2 different participants P4 and P5, respectively. (c) Overall performance for recovery participants in the test cohort with and without DWT, which calculates the optimal match between the predicted recovery trajectory and test results. (d) Projection of latent vectors learnt by the model for 3 different participants. The y axis from top to bottom indicates the test results over time. A clear change in each latent vector dimension transitioning from positive to negative can be observed (recovering user), and consistent and different patterns can be observed for COVID-19–positive and COVID-19–negative participants. (e,f) Scatter plot of symptoms vs probability of positive predictions for COVID-19–positive (e) and COVID-19–negative (f) participants, with a high correlation observed for COVID-19–positive participants and no correlation for COVID-19–negative participants. DWT; dynamic time warping.

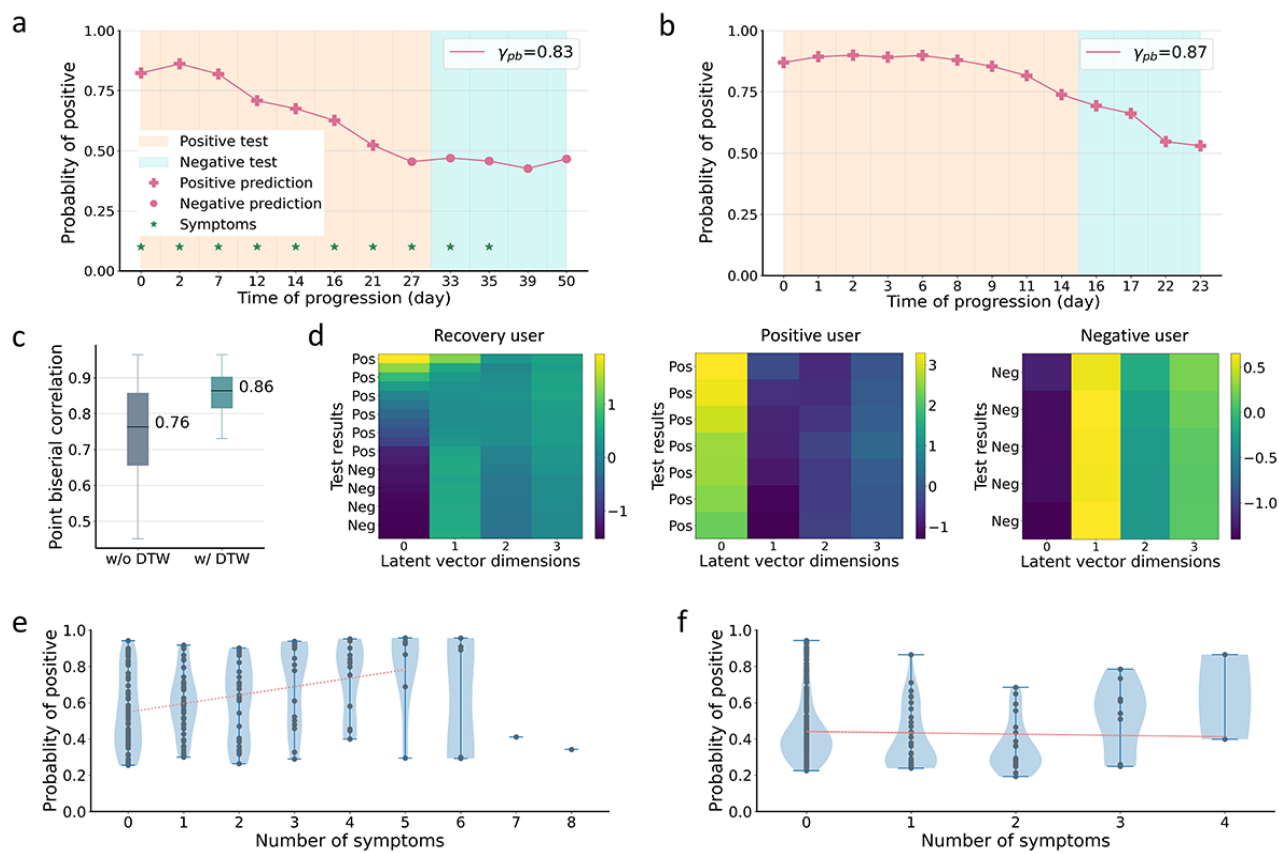

### Latent Space Visualization

Figure 7d shows the latent space visualization for 3 different participants. For the recovery user, a clear transition was observed for the first 3 dimensions when the participant recovered, while consistent but different patterns were observed for the COVID-19–positive and COVID-19–negative participants. This validates that the model can capture the different disease progression of different participants.

### Statistical Analysis

Figure 7e and Figure 7f display the correlations between the predicted probability and symptoms for COVID-19–positive and COVID-19–negative participants, respectively. With an increase in the number of symptoms, we observed a general increase in the predicted probability (Figure 7e). We further fit a line (red) for these participants, excluding those with more than 5 symptoms due to the limited number of samples. We observed a clear positive correlation. Conversely, for COVID-19–negative participants (Figure 7f), we observed no correlation between the probability and number of symptoms, suggesting that our model does not capture symptoms but information related to COVID-19.

Further analysis of the model predictions in the first 7 days was carried out on 12 (28.6%) eligible participants in the test cohort, where we found that 8 (66.7%) of 12 participants show an increase in predicted probability, similar to the statistical analysis (n=162, 65.3%). We hypothesized that the predicted progression in the first 7 days could provide a prompter indication of an individual's recovery rate.

## Discussion

### Principal Results

From a crowdsourced audio data set, we studied 212 longitudinal participants and developed a deep learning model for COVID-19 progression monitoring via audio signals. We showed that modeling audio dynamics longitudinally shows benefit for COVID-19 detection. Individual-level performance also displayed a significant improvement over baseline. The model capability to predict disease progression was validated. Successful tracking of reported test labels showed strong performance in disease progression with $\gamma_{pb}/\gamma=0.76$. We specifically focused on recovery prediction, with a correlation $\gamma_{pb}$ of 0.86 between the predicted progression trajectory and test results.

Individuals experience different disease progression trajectories, and our model can capture this variability among individuals. For the COVID-19–positive user P6 (Multimedia Appendix 3, Figure A3c), our model demonstrated a decrease in the predicted probability from day 0 to day 3, and this was followed by a slight increase from day 3 to day 6. For participant P7 (Multimedia Appendix 3, Figure A3d), our model predicted a continuous decrease from day 0 to 11. This suggests the effectiveness in predicting individual-specific disease progression trajectories. Though there is no reported severity of illness to validate the predicted probabilities, symptoms can be used as a reference. For participant P6, the number of symptoms increased from 3 to 6 at day 3 and decreased to 3 after. Therefore, it is reasonable to assume a worsening condition and an increase in the predicted probability.

Similarly, the model can also predict individual-specific recovery trajectories. A sharper recovery trend was observed for participant P1, with a 49.2% relative decrease in 21 days (Figure 6a) than for participants P4 (Figure 7a) and P5 (Figure 7b), with a 36.6% and 37.1% relative decrease in 21 and 22 days, respectively. This is consistent with the evidence that recovery tends to be faster in younger people [22], as participant P1 was in the age group of 20-29 years, while participants P4 and P5 were aged between 30-39 and 40-49 years, respectively. Though it is difficult to draw statistical conclusions due to the limited number of participants, these results still suggest differences in the predicted recovery rate for different individuals. In terms of practical applications, individual-specific recovery monitoring may be beneficial in providing prompt feedback to self-isolating patients and, more importantly, can provide treatment guidance for doctors according to each individual's recovery status. Specifically, when a sharp decrease in the predicted probability is observed, it indicates that the individual is recovering well. Conversely, no decrease in the predicted probability over a long period may require further or more effective treatment. Additionally, the predicted recovery trend could also be used to some extent for risk assessment of COVID-19 patients.

As our model showed strong performance in using longitudinal audio biomarkers, another important factor in the model deployment is the impact of sequence length, which was also analyzed to obtain insights into how many samples are needed for reliable predictions (Multimedia Appendix 5). The cumulative histogram suggests that the longer the sequence, the better the performance. For sequence lengths with more than 2 samples (Multimedia Appendix 5, Figure A2a) or around and more than 4 days (Multimedia Appendix 5, Figure A2b), the model can produce reasonably good predictions. For telemonitoring purposes, the use of audio recordings from the past 4 days would offer a more reliable prediction.

### Limitations

Our study also had several limitations. First, the testing cohort was relatively small with only 21 participants for the COVID-19–positive and COVID-19–negative groups. This may not comprehensively represent the target population. In addition, the self-reported test results may inevitably be noisy to some extent, where a mismatch between audio recordings and test results may exist. This is due to possible delays in participants reporting the test results. This mismatch introduced confounding variability into the model development that was not fully considered.

The other limitation in our study is the limited control over confounding factors. The age and gender groups were relatively balanced within and across the training, validation, and test sets, while language was only balanced between 3 partitions but still unbalanced within each data partition. Our model using a multitask framework mitigated the language impact, but some language bias might remain due to the limited number of samples of some language subgroups.

We also acknowledged that changes in voice may be attributed to not only COVID-19 infection but also other factors (eg, mental state or other respiratory infections, such as influenza). To validate whether the model captures changes caused by COVID-19 instead of other factors, a large amount of longitudinal data that have the corresponding labels (eg, emotion state, influenza) is required to develop and evaluate the model. Collecting such data is difficult and time-consuming, which is our long-term goal.

It is also worth noting that the predicted disease progression trend matches the test results, but for some users, probabilities may be overall high or low over the course of COVID-19 progression. This suggests individual differences in audio characteristics. Though our model resolves this better than simple sample-based models by capturing past audio signals, it is a universal model and therefore still imprecise. The development of participant-specific models is on our future agenda, but more data needs to be collected for this purpose.

### Conclusion

In conclusion, by modeling audio biomarkers longitudinally with sequential machine learning techniques, we proposed audio-based diagnostics with longitudinal data as a robust technique for COVID-19 progression tracking. We showed that our system is able to monitor disease progression, especially the recovery trajectory of individuals. This work not only provides a flexible, affordable, and timely tool for COVID-19 tracking but also provides a proof of concept of how telemonitoring could be applicable to respiratory diseases monitoring in general.

### Acknowledgments

This work was supported by European Research Council (ERC) Project 833296 (EAR). We thank everyone who volunteered their data.

## Data Availability

The data were sensitive as voice sounds can be deanonymized. Anonymized data will be made available for academic research upon requests directed to the corresponding author. Institutions need to sign a data transfer agreement with the University of Cambridge to obtain the data. A copy of the data will be transferred to the institution requesting the data. We already have the data transfer agreement in place. Python code and parameters used for training of neural networks will be available on GitHub for reproducibility purposes.

## Authors' Contributions

AF, CM, and PC designed the study. AH, AG, CS-B, DS, and JC designed and implemented the mobile app to collect the sample data. AG designed and implemented the server infrastructure. JH, TD, and TX selected the data for analysis. DS, TD, and TX developed the neural network models. TD conducted the experiments, performed the statistical analysis, wrote the main draft of the manuscript, and generated tables and figures. JH and TX cowrote the manuscript. All authors vouch for the data, analyses, and interpretations. All authors critically reviewed, contributed to the preparation of the manuscript, and approved the final version.

## Conflicts of Interest

None declared.

## Multimedia Appendix 1

Details of model development and validation, including data collection, data selection, data augmentation, model architecture, and model training and evaluation.
[DOCX File , 25 KB-Multimedia Appendix 1]

## Multimedia Appendix 2

Data statistics in terms of gender, age, and language.
[DOCX File , 574 KB-Multimedia Appendix 2]

## Multimedia Appendix 3

Additional examples of strong disease progression predictions.
[DOCX File , 242 KB-Multimedia Appendix 3]

## Multimedia Appendix 4

Additional examples of weak disease progression predictions.
[DOCX File , 419 KB-Multimedia Appendix 4]

## Multimedia Appendix 5

Detailed analysis on the impact of sequence length for COVID-19 detection.
[DOCX File , 957 KB-Multimedia Appendix 5]

## References

1. Vogels CBF, Brito AF, Wyllie AL, Fauver JR, Ott IM, Kalinich CC, et al. Analytical sensitivity and efficiency comparisons of SARS-CoV-2 RT-qPCR primer-probe sets. Nat Microbiol 2020 Oct;5(10):1299-1305. [doi: 10.1038/s41564-020-0761-6] [Medline: 32651556]
2. Cevik M, Kuppalli K, Kindrachuk J, Peiris M. Virology, transmission, and pathogenesis of SARS-CoV-2. BMJ 2020 Oct 23;371:m3862. [doi: 10.1136/bmj.m3862] [Medline: 33097561]
3. Fan L, Liu S. CT and COVID-19: Chinese experience and recommendations concerning detection, staging and follow-up. Eur Radiol 2020 May 06;30(9):5214-5216. [doi: 10.1007/s00330-020-06898-3]
4. Deshpande G, Batliner A, Schuller BW. AI-Based human audio processing for COVID-19: a comprehensive overview. Pattern Recognit 2022 Feb;122:108289. [doi: 10.1016/j.patcog.2021.108289] [Medline: 34483372]
5. Ates HC, Yetisen AK, Güder F, Dincer C. Wearable devices for the detection of COVID-19. Nat Electron 2021 Jan 25;4(1):13-14. [doi: 10.1038/s41928-020-00533-1]
6. Channa A, Popescu N, Skibinska J, Burget R. The rise of wearable devices during the COVID-19 pandemic: a systematic review. Sensors (Basel) 2021 Aug 28;21(17):5787 [FREE Full text] [doi: 10.3390/s21175787] [Medline: 34502679]
7. Barr PJ, Ryan J, Jacobson NC. Precision assessment of COVID-19 phenotypes using large-scale clinic visit audio recordings: harnessing the power of patient voice. J Med Internet Res 2021 Feb 19;23(2):e20545. [doi: 10.2196/20545]

XSL·FO
RenderX

8. Stasak B, Huang Z, Razavi S, Joachim D, Epps J. Automatic detection of COVID-19 based on short-duration acoustic smartphone speech analysis. J Healthc Inform Res 2021 Mar 11;5(2):201-217. [doi: 10.1007/s41666-020-00090-4] [Medline: 33723525]
9. Miranda ID, Diacon AH, Niesler NR. A comparative study of features for acoustic cough detection using deep architectures. 2019 Presented at: 41st Annual International Conference of the IEEE Engineering in Medicine and Biology Society (EMBC); July 23-27, 2019; Berlin, Germany p. 2601-2605. [doi: 10.1109/embc.2019.8856412]
10. Al Ismail M, Deshmukh S, Singh R. Detection of COVID-19 through the analysis of vocal fold oscillations. 2021 Presented at: 2021-2021 IEEE International Conference on Acoustics, Speech and Signal Processing (ICASSP); June 6-11, 2021; Toronto, Ontario, Canada p. 1035-1039. [doi: 10.1109/icassp39728.2021.9414201]
11. Deshmukh S, Al Ismail M, Singh R. Interpreting glottal flow dynamics for detecting COVID-19 from voice. 2021 Presented at: 2021-2021 IEEE International Conference on Acoustics, Speech and Signal Processing (ICASSP); June 6-11, 2021; Toronto, Ontario, Canada p. 1055-1059. [doi: 10.1109/icassp39728.2021.9414530]
12. Deshpande G, Schuller BW. Audio, Speech, Language, and Signal Processing for COVID-19: A Comprehensive Overview. arXiv Preprint posted online November 29, 2020. [doi: 10.48550/arXiv.2011.14445]
13. Laguarta J, Hueto F, Subirana B. COVID-19 artificial intelligence diagnosis using only cough recordings. IEEE Open J Eng Med Biol 2020;1:275-281. [doi: 10.1109/ojemb.2020.3026928]
14. Imran A, Posokhova I, Qureshi HN, Masood U, Riaz MS, Ali K, et al. AI4COVID-19: AI enabled preliminary diagnosis for COVID-19 from cough samples via an app. Inform Med Unlocked 2020;20:100378. [doi: 10.1016/j.imu.2020.100378] [Medline: 32839734]
15. Brown C, Chauhan J, Grammenos A, Han J, Hasthanasombat A, Spathis D, et al. Exploring automatic diagnosis of COVID-19 from crowdsourced respiratory sound data. arXiv Preprint posted online June 10, 2020. [doi: 10.48550/arXiv.2006.05919]
16. Pinkas G, Karny Y, Malachi A, Barkai G, Bachar G, Aharonson V. SARS-CoV-2 detection from voice. IEEE Open J Eng Med Biol 2020;1:268-274. [doi: 10.1109/ojemb.2020.3026468]
17. Han J, Brown C, Chauhan J, Grammenos A, Hasthanasombat A, Spathis D, et al. Exploring automatic COVID-19 diagnosis via voice and symptoms from crowdsourced Data. 2021 Presented at: 2021-2021 IEEE International Conference on Acoustics, Speech and Signal Processing (ICASSP); June 6-11, 2021; Toronto, Ontario, Canada p. 8328-8332. [doi: 10.1109/ICASSP39728.2021.9414576]
18. Coppock H, Gaskell A, Tzirakis P, Baird A, Jones L, Schuller B. End-to-end convolutional neural network enables COVID-19 detection from breath and cough audio: a pilot study. BMJ Innov 2021 Apr 16;7(2):356-362 [FREE Full text] [doi: 10.1136/bmjinnov-2021-000668] [Medline: 34192022]
19. Andreu-Perez J, Perez-Espinosa H, Timonet E, Kiani M, Giron-Perez M, Benitez-Trinidad AB, et al. A generic deep learning based cough analysis system from clinically validated samples for point-of-need Covid-19 test and severity levels. IEEE Trans Serv Comput 2021:1-1. [doi: 10.1109/TSC.2021.3061402]
20. Yu F, Yan L, Wang N, Yang S, Wang L, Tang Y, et al. Quantitative detection and viral load analysis of SARS-CoV-2 in infected patients. Clin Infect Dis 2020 Jul 28;71(15):793-798 [FREE Full text] [doi: 10.1093/cid/ciaa345] [Medline: 32221523]
21. Wu J, Li W, Shi X, Chen Z, Jiang B, Liu J, et al. Early antiviral treatment contributes to alleviate the severity and improve the prognosis of patients with novel coronavirus disease (COVID-19). J Intern Med 2020 Jul 20;288(1):128-138. [doi: 10.1111/joim.13063] [Medline: 32220033]
22. Voinsky I, Baristaite G, Gurwitz D. Effects of age and sex on recovery from COVID-19: analysis of 5769 Israeli patients. J Infect 2020 Aug;81(2):e102-e103 [FREE Full text] [doi: 10.1016/j.jinf.2020.05.026] [Medline: 32425274]
23. Lechien JR, Chiesa-Estomba CM, Place S, Van Laethem Y, Cabaraux P, Mat Q, COVID-19 Task Force of YO-IFOS. Clinical and epidemiological characteristics of 1420 European patients with mild-to-moderate coronavirus disease 2019. J Intern Med 2020 Sep;288(3):335-344 [FREE Full text] [doi: 10.1111/joim.13089] [Medline: 32352202]
24. Bi Q, Wu Y, Mei S, Ye C, Zou X, Zhang Z, et al. Epidemiology and transmission of COVID-19 in 391 cases and 1286 of their close contacts in Shenzhen, China: a retrospective cohort study. Lancet Infect Dis 2020 Aug;20(8):911-919. [doi: 10.1016/s1473-3099(20)30287-5]
25. Siddiqi HK, Mehra MR. COVID-19 illness in native and immunosuppressed states: a clinical-therapeutic staging proposal. J Heart Lung Transplant 2020 May;39(5):405-407 [FREE Full text] [doi: 10.1016/j.healun.2020.03.012] [Medline: 32362390]
26. Chen J, Qi T, Liu L, Ling Y, Qian Z, Li T, et al. Clinical progression of patients with COVID-19 in Shanghai, China. J Infect 2020 May;80(5):e1-e6 [FREE Full text] [doi: 10.1016/j.jinf.2020.03.004] [Medline: 32171869]
27. University of Cambridge. COVID-19 Sounds App. URL: https://www.covid-19-sounds.org/en/ [accessed 2022-06-06]
28. Xia T, Spathis D, Brown C, Ch J, Grammenos A, Han J, et al. COVID-19 sounds: a large-scale audio dataset for digital respiratory screening. 2021 Presented at: Thirty-fifth Conference on Neural Information Processing Systems Datasets and Benchmarks Track (Round 2); December 6-14, 2021; Virtual-only.
29. Hershey S, Chaudhuri S, Ellis DPW, Gemmeke JF, Jansen A, Moore RC, et al. CNN architectures for large-scale audio classification. 2017 Presented at: 2017 IEEE International Conference on Acoustics, Speech and Signal Processing (ICASSP); June 19, 2017; New Orleans, LA p. 131-135. [doi: 10.1109/icassp.2017.7952132]

**Abbreviations**

**AUROC:** area under the receiver operating characteristic curve
**CT:** computed tomography
**DTW:** dynamic time warping
**GRU:** gated recurrent unit
**RT-PCR:** reverse transcription polymerase chain reaction